\documentclass[showpacs,amsmath,amssymb,twocolumn]{revtex4}

\usepackage{graphicx}
\usepackage{dcolumn}
\usepackage{fancyhdr}
\usepackage{indentfirst}
\usepackage{newlfont}
\usepackage{amssymb}
\usepackage{amsmath}
\usepackage{latexsym}
\usepackage{amsthm}

\begin{document}

\title{"High pressure behavior of Ga-doped LaMnO$_{3}$: a combined X-ray diffraction and optical spectroscopy study."}
\author{M. Baldini$^{1}$, L.Malavasi$^{2}$, D. Di Castro$^{1}$, A. Nucara$^{1}$ W. Crichton$^{3}$, M. Mezouar$^{3}$, J. Blasco$^{4}$, P. Postorino$^{1}$}
\affiliation{$^{1}$ INFM and Dipartimento di Fisica, Universit\`a
"Sapienza", P.le A. Moro 4, I-00187 Roma, Italy.}
\affiliation{$^{2}$ Dipartimento di chimica "M.Rolla", INSTM and
IENI-CNR, Universit\`a di Pavia, Viale Taramelli 16, I-27100
Pavia, Italy} \affiliation{$^{3}$European Synchrotron Radiation
Facility, BP 220, F-38043 Grenoble, France}
\affiliation{$^{4}$Instituto de Ciencia de Materiales de
Arag\`{o}n, Departamento de Fisica de la Materia Condensada, CSIC-
Universidad de Zaragoza, Pedro Cerbuna 12, 50009 Zaragoza, Spain}
\begin{abstract}
The pressure effects on the \emph{JT} distortion of three
representative compounds belonging to the LaMn$_{1-x}$Ga$_x$O$_3$
(x= 0.2, 0.3, 0.4) family was widely investigated by means of
X-ray diffraction and Raman spectroscopy. A compound with a fully
\emph{JT} distorted structure (x=0.2), one with regular octahedra
(x=0.6) and one in an intermediate configuration (x=0.3) were
selected. A pressure induced transitions from the orthorhombic
\emph{Pbnm} phase towards structures with higher symmetry were
observed in all the samples. Both Raman and X-ray data confirm
that the most important structural effect of pressure is that of
reducing the octahedral distortion. The appearance of a feature in
the lattice parameter behavior connected to a structural
instability was also detected, pointing out the key role of the
\emph{JT} distortion in stabilizing the manganite structures. On
the other hand, the complete suppression of the \emph{JT}
distortion in the high-pressure phases cannot be claimed. The
Raman spectra collected from more distorted compounds (x=0.2, 0.3)
reveal clearly the coexistence of domains of distorted and more
regular octahedra in a certain pressure range. The first sketch of
the Pressure vs. Ga-content phase diagram was drawn.
\end{abstract}

\maketitle

\section{Introduction}
In recent years large efforts have been devoted to the
comprehension of the fundamental mechanisms involved in the charge
doped colossal magnetoresistance manganites \cite{Cheong}.
Theoretical and experimental results show the close correlation
between structural, orbital, and electronic degrees of freedom,
which have to be taken into account to shed light on the complex
physics underlying the unique properties of these compounds. In
particular, the orthorhombic parent compound LaMnO$_3$ is
considered as the prototype of a \emph{coherent} Jahn-Teller
(\emph{JT}) system, where a cooperative tetragonal deformation of
the MnO$_6$ octahedra occurs \cite{Good100}. The LaMnO$_3$
structure can be viewed as a sequence of alternate short and long
Mn-O bonds in the \emph{ab} plane \cite{dtype}. Furthermore, the
occurrence of \emph{d}-type orbital ordered state
\cite{OO,mizokawa}, which plays a fundamental role in stabilizing
the anisotropic A-type antiferromagnetic state, has also been
clearly pointed out \cite{Good100,Hennion,Kim}. The replacement of
trivalent La with a divalent ion (Ca, Sr) corresponds to an
effective hole-doping. In the $0.2 < x < 0.5$ concentration range,
colossal magnetoresistance \cite{cmr} and a ferromagnetic metallic
ground state are observed, which are qualitatively well explained
by means of the double exchange (DE) mechanism \cite{DE}. Although
the above picture appears to be rather well established, recent
investigations carried out on Ga$^{3+}$ doped compounds have
opened up new interesting, and still unsolved, questions. Indeed,
the replacement of Mn$^{3+}$ by a nonmagnetic and non Jahn-Teller
(\emph{JT}) ion like Ga$^{3+}$, surprisingly induces a
ferromagnetic insulating ground state in the
LaMn$_{1-x}$Ga$_x$O$_3$ compounds for
$0.4\lesssim$\emph{x}$\leq0.6$ \cite{neutroni}. It is worth
noticing that the absence of hole-doping in these compounds does
not allow to ascribe this singular behavior to a DE mechanism. All
the structural studies on this family of compounds
\cite{neutroni,blasco1,blasco2} show that the Ga substitution
progressively reduces the \emph{JT} distortion leading, for
\emph{x} $\geq 0.6$, to a crystallographic structure with all the
MnO$_6$ octahedra in a symmetric configuration. The behavior of
magnetization can be described by a spin-flipping model: this
theoretical result seems to suggest that the Ga doping, by
removing the \emph{JT} distortion, perturbs the orbital order and
favors the ferromagnetic ground state \cite{Gehring}. On the other
hand, Raman and X-ray diffraction experiments previously performed
on the parent compound have suggested that the cooperative
\emph{JT} effect in LaMnO$_3$ \cite{loa} can be reduced by
application of hydrostatic pressure. Therefore, hydrostatic
pressure, coupled with Raman spectroscopy and X-ray diffraction,
turns out to be a very effective tool to investigate the role of
the \emph{JT} distortion \cite{loa} in the Ga doped LaMnO$_{3}$. A
recent Raman experiment \cite{noigallati}, performed on several
compounds of the LaMn$_{1-x}$Ga$_x$O$_3$ family
($0.1<$\emph{x}$\leq0.8$) confirms the removal of the \emph{JT}
octahedra distortion on increasing the Ga doping. Moreover, high
pressure Raman measurements performed on the \emph{x}$=0.2$ (LG20)
and  \emph{x}$=0.6$ (LG60) samples have revealed that pressure, as
well as doping, favors the symmetrization process
\cite{noigallati}. However, differences between the doping and
pressure effects have been also pointed out by these Raman
measurements \cite{noigallati}.

In the present paper, a deeper and more extended investigation of
the pressure effects on the \emph{JT} distortion \cite{loa} in the
Ga doped LaMnO$_{3}$ has been carried out. In particular high
pressure X-ray diffraction measurements were carried out over the
$0-18$ GPa pressure range on three Ga doped manganites at room
temperature, namely the \emph{x}$=0.2$ (LG20), \emph{x}$=0.3$
(LG30) and \emph{x}$=0.6$ (LG60) compounds. A new high-pressure
Raman measurement ($0-13$ GPa) has been also carried out on the
LG30 sample in order to have, together with the data reported in
Ref. \onlinecite{noigallati}, a complete set of high-pressure
Raman and  X-ray diffraction data for the LG20, LG30, and LG60
samples. These compositions have been chosen since a fully
\emph{JT} distorted structure is observed in the LG20 sample,
opposite to the symmetric regular octahedra observed in the LG60
compound, whereas the LG30 sample is in an intermediate
configuration. Both the experimental techniques show that
hydrostatic pressure is an effective tool in order to reduce the
\emph{JT} distortion. The pressure induced reduction of \emph{JT}
distortion is clearly observed by both Raman and X-ray
measurements. On the other hand, the pressure induced
symmetrization process turn out to be different from that induced
by doping.
\section{Experimental}
The LaMn$_{1-x}$Ga$_{x}$O$_{3}$ samples (LG20, LG30 and LG60) were
prepared following a conventional ceramic procedure
\cite{neutroni}. Stoichiometric amounts of La$_{2}$O$_{3}$,
MnCO$_{3}$, and Ga$_{2}$O$_{3}$ were mixed, milled, and fired at
1250 C during 24 h in air. Then, the resulting powders were
milled, pressed into pellets and sintered at 1250 °C for another
24 h in air. Finally, the pellets were milled, repressed and
sintered at 1400 $^{\circ}C$ for 48 h in an argon atmosphere. The
oxygen content of the samples was analyzed using standard redox
titration with KMnO$_{4}$ and Mohr' salt. These samples were
oxygen-stoichiometric, within the analysis accuracy $\pm 0.02$
\cite{neutroni}.\\High pressure X-ray diffraction data were
collected on the ID27 beamline at the ESRF Facility in Grenoble,
which is dedicated at high pressure diffraction experiments. The
beam ($\lambda$ = 0.41 $\AA$) size on the sample is normally about
30 x 30 $\mu$m$^{2}$. A membrane diamond anvil cell (MDAC)
equipped with low fluorescence IIA, 400 $\mu$m culet diamonds has
been used. The samples were finely milled and an high-pressure
MDAC-loading with N$_{2}$, as pressure transmitting medium was
performed. The image plate detector is a Mar345 reader. The
experimental line width in the diffraction pattern is smaller than
$\Delta \vartheta = 0.05°$ FWHM (Full Width at Half Maximum). The
X-ray data were refined by means of the Fullprofile software
\cite{fullprof}.

Raman spectra were measured in back-scattering geometry, using a
micro-Raman spectrometer (LABRAM by Jobin Ivon) with a
charge-coupled device (CCD) detector and an adjustable notch
filter. The samples were excited by the $632.8$ nm line of a $15$
mW He-Ne Laser. The confocal microscope was equipped with a $20$X
magnification objective which gives laser spot about $5 \mu m^2$
wide at the sample surface. The Raman spectra were collected in
the $200$-$1100$ cm$^{-1}$ frequency range. The MDAC was equipped
with low fluorescence IIA diamonds with $800$ $\mu$m culet
diameter. The hydrostatic conditions for the sample were ensured
by the NaCl salt loaded with the sample. Moreover, owing to the
high thermal conductivity of diamond, spurious effects due to
laser-induced sample heating can be prevented \cite{dho}. Since
the laser spot size ($5$ $\mu$m$^2$) is much smaller than the
sample diameter size ($200$ $\mu$m$^2$) we were able to collect
Raman spectra from three different points of the sample. A good
spectral reproducibility of the spectra was obtained at each
pressure. The spectra from different zones were analyzed
separately using a DHO function \cite{dho} and the best fit
parameters were averaged. The pressure measurement is obtained, in
both the experimental techniques, exploiting the standard ruby
fluorescence technique \cite{ruby}.
\section{Experimental results and discussion}

The LaMn$_{1-x}$Ga$_{x}$O$_{3}$ compounds belong to the family of
oxides with a perovskite-like structure. At ambient pressure the
samples show an orthorhombic \emph{Pbnm} structure. Two
orthorhombic \emph{Pbnm} phases can be distinguished: the first,
usually named O', is observed for \emph{x}$\leq$0.5 and it is
related to a \emph{JT} \emph{coherent} deformation of the
MnO$_{6}$ octahedra and to orbital ordered state; the other one,
named O, is observed for \emph{x}$>0.5$ and it is related to
regular octahedra and to an orbital disordered phase
\cite{neutroni, blasco1,blasco2}. The three presently investigated
samples thus belong to both the phases and namely LG20 and LG30 to
the O' phase, and LG60 to the O one.

\subsection{High pressure X-ray diffraction}
\begin{figure}
  \includegraphics[width=8cm]{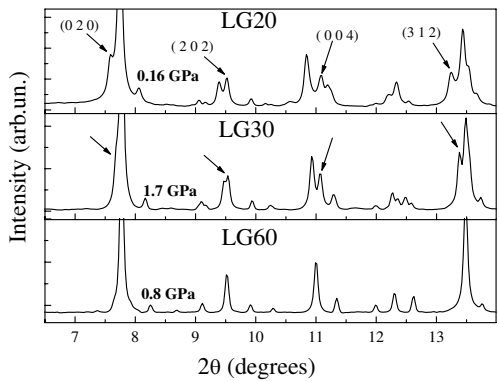}\\
  \caption{X-ray diffraction patterns collected at the lowest pressure.
 The black arrows indicate the (0 2 0), (2 0 2), (0 0 4) and the (3 1 2) reflections which are
 the marker of the structural transition from the O' to the O phase.}\label{FIG1}
\end{figure}

\begin{figure}[h]
  \includegraphics[width=9 cm]{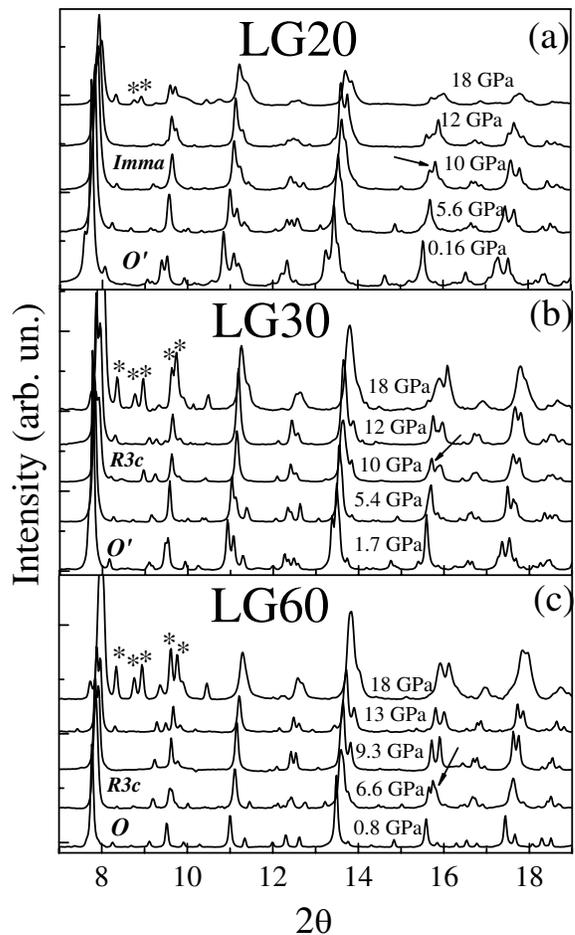}\\
  \caption{Diffraction patterns of LG20 (a), LG30 (b) and LG60 (c) samples at selected pressure.
  The black arrow indicates the peak splitting occurring at the structural
  transition. The asterisks mark the reflections due to the
nitrogen pressure medium in the $\epsilon$-phase.}\label{FIG2}
\end{figure}

\begin{figure}[h!]
  \includegraphics[width=8 cm]{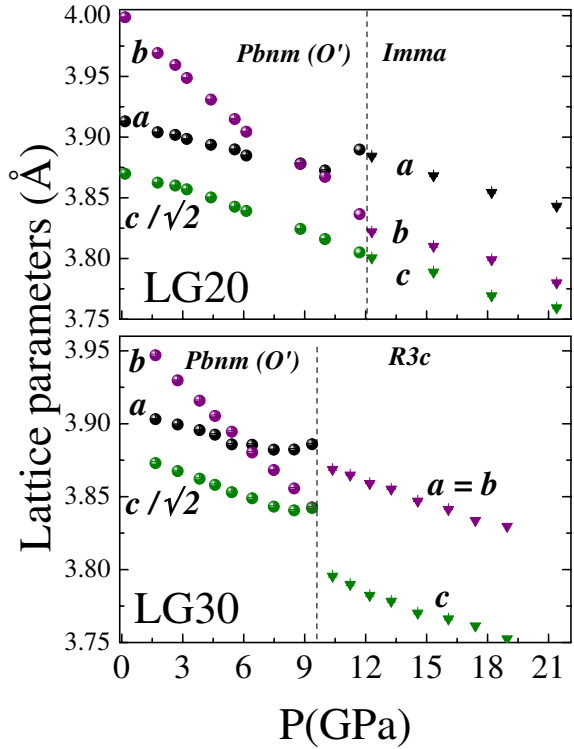}\\
  \caption{Pressure dependence of the LG20 and LG30 lattice parameters. Vertical dashed lines mark the
transition pressure.}\label{FIG3}
\end{figure}

\begin{figure}[h!]
  \includegraphics[width=9 cm]{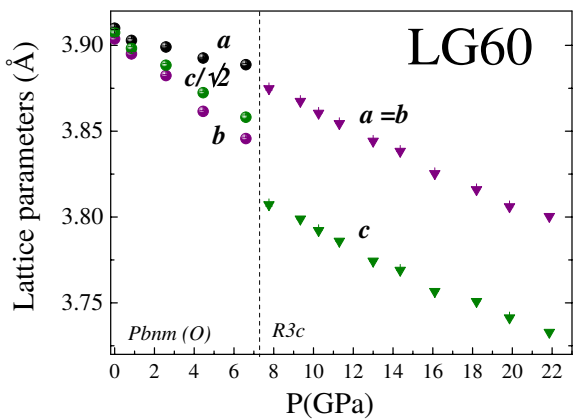}\\
  \caption{Pressure dependence of the LG60 lattice parameters.}\label{FIG4}
\end{figure}

Figure \ref{FIG1} reports the X-ray diffraction patterns collected
on the LG20, LG30 and LG60 samples at the lowest pressure. The
Rietveld refinements performed on these patterns are in good
agreement with the O' (LG20 and LG30) and O (LG60) \emph{Pbnm}
phases reported in literature \cite{neutroni}. The disappearance
of the (0 2 0), (2 0 2), (0 0 4) and the (3 1 2) reflections with
doping (see Fig.\ref{FIG1}) marks the structural transition from
the \emph{JT} distorted O' to the undistorted O \emph{Pbnm} phase,
previously reported in ref.
\cite{neutroni}.\\
In Fig.\ref{FIG2} the diffraction patterns of the three samples
collected at selected pressures are shown.
In all the measured samples, the lattice compression induces a
clear shift of the diffraction peaks towards higher angles as the
pressure is increased. At a first glance, the appearance of new
diffraction peaks is observed in the LG20, LG30 and LG60
diffraction patterns collected at the highest pressure, which can
be associated to a a pressure induced structural transitions. The
clear peak splitting of the reflection centered at around
15$^{\circ}$ (see the black arrow in Fig.\ref{FIG2} (a), (b) and
(c)), marks well the onset of the structural transition in all the
samples.\\All the patterns were analyzed with the Rietveld method.
The change in the peak intensities together with the peak
spreading, both induced by pressure, do not allow to extract
trustworthy information from the atomic position refinements, thus
only the pressure dependence of the lattice parameters, is
reported in this paper. The results are shown in Fig.\ref{FIG3}
(LG20 and LG30) and in Fig.\ref{FIG4} (LG60). As the pressure is
increased, structural transitions are observed in all the measured
compounds. A transition from an orthorhombic \emph{Pbnm} O'-type
to an orthorhombic \emph{Imma} phase occurs at around 12 GPa in
the LG20 sample, and a transition from orthorhombic to
rhombohedral \emph{R$\overline{3}$c} phase is observed at around
10 and 7 GPa in the LG30 and LG60 compounds, respectively. The
onset of the higher symmetric orthorhombic \emph{Imma} and the
rhombohedral \emph{R$\overline{3}$c} space groups indicates that
pressure really promotes the octahedra symmetrization. However
more detailed information can be deduced by the lattice parameter
pressure behavior. It is worth to notice that the onset of
\emph{coherent} \emph{JT} distortion in the \emph{Pbnm} symmetry
lowers the \emph{a}/$ \frac{\emph{c}}{\sqrt{2}}$ ratio and yields
\emph{c}/$\sqrt{2}$ $<$ \emph{a} $<$ \emph{b} \cite{landolt}.
Therefore, in the orthorhombic structure \emph{Pbnm}, if the
\emph{c} parameter is smaller than \emph{a} and \emph{b}, the
occurrence of an orbital ordering can be hypothesized
\cite{Woodward}. Moreover, since the \emph{JT} distortion mainly
affects the Mn-O distance in the \emph{ab} plane, changes in the
\emph{b} cell parameter can be a good indicator of the magnitude
of the cooperative \emph{JT} distortion \cite{Woodward}.
Within the O' phase both LG20 and LG30 samples show the \emph{b}
pressure rate of compression larger than that of the \emph{a} and
\emph{c} parameters so that a crossing of the \emph{a} and
\emph{b} lattice parameters occurs at around 8 GPa (LG20) and 6
GPa (LG30). This finding is once more compatible with a
pressure-induced reduction of the \emph{JT} distortion in both the
samples. Above the transition pressure (P= 12 GPa) three distinct
lattice parameters are still observed in LG20 (\emph{Imma} phase)
and the values of the \emph{c} parameter keep smaller than those
of the \emph{b} and the \emph{a} parameters. These findings thus
suggest, that the pressure-induced removal of \emph{JT} is not
completed and that the persistence of a residual distortion also
in the \emph{Imma} high pressure phase may be conjectured. On the
other hand, the occurrence of only two Mn-O distances in the
\emph{Imma} structure instead of the three ones characteristic of
the O' phase is an indication of the loss of \emph{coherence} of
the octahedral deformation.\\Above about 10 GPa LG30 enters the
\emph{R$\overline{3}$c} phase where the occurrence of
\emph{coherent} \emph{JT} distortion is not allowed owing to the
rhombohedral tilting \cite{Woodward}. Nevertheless the presence of
\emph{dynamic} or static \emph{incoherent} \emph{JT} distortions,
previously claimed in several \emph{R$\overline{3}$c} manganite
compounds \cite{radaelli}, cannot be ruled out. The data analysis
of the two sample (LG20 and LG30) thus coherently indicates that
the reduction of the \emph{JT} distortion is promoted by
hydrostatic pressure although a complete
removal of octahedral distortion cannot be claimed.\\
At variance from the LG20 and LG30, the lattice parameters of LG60
are almost coincident at ambient pressure (see Fig.\ref{FIG4}) and
show a peculiar pressure behavior. This is not surprising since
LG60 at ambient pressure shows undistorted octahedra within the
orthorhombic \emph{Pbnm} O-type (\emph{a}$>$\emph{b}$\simeq$
\emph{c}/$\sqrt{2}$) phase, whereas both LG20 and LG30 are
characterized by \emph{JT} distorted within the \emph{Pbnm}
O'-type phase. On applying pressure, the three LG60 lattice
parameters get more and more distinct up to 7 GPa when a
transition to the \emph{R$\overline{3}$c} phase occurs and only
two lattice parameters are necessary to fit the diffraction data.
We notice that although the values of lattice parameters, actually
identical at ambient pressure distinguish each other as the
pressure is increased over the 0-7 GPa pressure range, the ratio
between the \emph{a}, \emph{b}, \emph{c} lattice parameters
suggests the absence of \emph{JT} distortion until the \emph{Pbnm}
is preserved \cite{landolt}. In the high pressure regime (P$>$7
GPa) the \emph{R$\overline{3}$c} phase does not allow the presence
of \emph{coherent} JT distortions as mentioned above, but in this
case where the compound shows regular octahedra even at ambient
pressure, the presence of \emph{dynamic} or static
\emph{incoherent} \emph{JT} distortions is very unlikely.

The crossing of the \emph{b} and \emph{a} lattice parameters,
observed in both LG30 (at P=6 GPa) and LG20 (at P=8 GPa) deserves
further discussion. This feature has been already observed in the
pressure dependence of LaMnO$_{3}$ lattice parameters at around 10
GPa \cite{loa}, and previously observed in the evolution of
several AMnO$_{3}$ manganites  lattice parameters as a function of
the rare earth (A) ionic radius \cite{good,Woodward}, i.e., as a
function of the chemical pressure. The crossover of \emph{b} and
\emph{a}, as the ionic radius increases above 1.11 $\AA$, points
out a structural instability in the orthorhombic \emph{Pbnm} O
symmetry which leads to a structural transition from orthorhombic
to rhombohedral (\emph{R$\overline{3}$c}) phase \cite{good}. It
seems to be a general phenomenon, occurring during the
symmetrization process, that systematically prevents a continuous
evolution from the orthorhombic to the cubic AMnO$_{3}$ perovskite
structure \cite{good}. The \emph{a}-\emph{b} crossing observed in
the Ga doped compounds and in LaMnO$_{3}$ \cite{loa}, as the
pressure is increased, could be ascribed to the same process.
First of all this crossing is a clear mark of the octahedra
symmetrization and the whole of  our results are consistent with
this statement. Indeed, by comparing the pressure behavior of the
lattice parameters of the distorted LG20 and LG30 samples, the
crossing occurs at a pressure threshold which is higher in LG20
(P=8 GPa) than in the less distorted LG30 (P=6 GPa) which shows
weaker \emph{JT} distortion and closer \emph{a} and \emph{b}
parameters values. Consistently the \emph{a}-\emph{b} crossing
does not occur in LG60 where the \emph{JT} distortion is already
removed and regular octahedra are already observed at ambient
pressure. Moreover this result not only points out the connection
between this feature and the reduction of \emph{JT} distortion but
it also gives a strong indication of the a structural instability
occurring in the system. This instability is the precursor of a
structural transition and prevents the direct evolution from a
distorted to an undistorted cubic structure. Therefore, the
pressure induced symmetrization process appears to occur step by
step, through subsequent transitions to higher symmetry phases.

Some information on the orbital order state can be also indirectly
deduced from the lattice parameter behavior. As mentioned above,
in the orthorhombic structure \emph{Pbnm}, if the \emph{c}
parameter is smaller than \emph{a} and \emph{b}, the occurrence of
an orbital ordering can be hypothesized \cite{Woodward}.
Therefore, the occurrence of an orbitally ordered state can be
reasonably supposed as long as the system preserves the
orthorhombic O' \emph{Pbnm} phase in the LG20 and LG30 samples
(see Fig.\ref{FIG3}). As to LG20 a clear understanding relative to
the existence of an orbitally ordered state above the structural
transitions is far to be completely achieved. In this case the
\emph{c} behavior above 12 GPa suggests the existence of some kind
of orbital orientations in the LG20 \emph{Imma} high pressure
phase. Similar problem was discussed for the LaMnO$_{3}$ in the
high pressure regime (P$>$ 15 GPa) \cite{trimarchi}. A mixture of
\emph{a}-type and \emph{d}-type \emph{JT} distortion and thus of
\emph{a} and \emph{d}-type orbital configurations was supposed in
the LaMnO$_{3}$ high pressure phase \cite{trimarchi}. This
hypothesis is consistent with the results obtained in the LG20
sample. Indeed, it requires the loss of the \emph{JT} distortion
\emph{coherence}, which results in the \emph{Imma} phase. However,
the appearance of a specific orbital order configuration cannot be
inferred and the hypothesis of an orbital disordered state cannot
be ruled out.\\The existence of an orbital ordered state is not
allowed in the high pressure \emph{R$_{3}$c} phase observed in
both LG30 and LG60. Therefore, in the LG30 compound the occurrence
of the \emph{d}-type orbitally ordered state can be supposed only
below 10 GPa. On the contrary, in the LG60 sample the occurrence
of an orbitally ordered state can be also certainly ruled out over
the whole pressure range. Indeed, \emph{c} is larger than \emph{b}
in the \emph{Pbnm} O phase (see Fig.\ref{FIG4}), revealing an
orbitally disordered state.

\begin{figure}[h!]
  \includegraphics[width=9 cm]{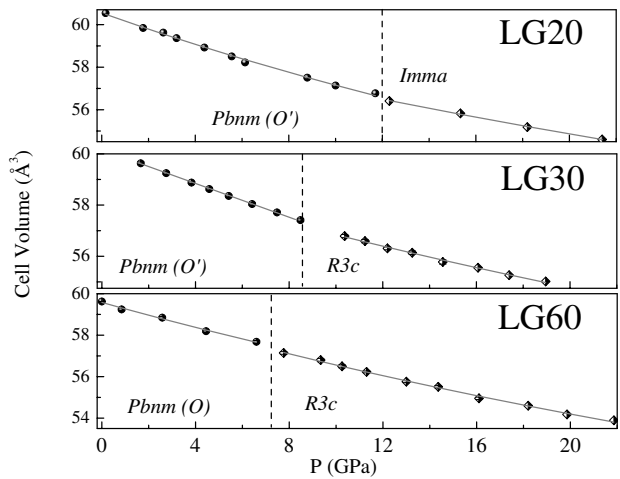}\\
  \caption{Pressure dependence of the cell volumes. The experimental data
(circles) was fitted with a Birch-Murnaghan curve (black line).
}\label{FIG5}
\end{figure}

The pressure dependence of the cell volumes was calculated for
each samples, and shown in Fig.\ref{FIG5}.
The cell volume behavior as a function of pressure is well
described by a Birch-Murnaghan curve \cite{birk1,birk2}, which
correlates the pressure to V/V$_{0}$ (V$_{0}$ is the cell volume
at ambient pressure) as follows:
\begin{equation}\label{birch}
   P = \frac{K_{0}}{K_{0}^{'}}\left[\left(\frac{V_{0}}{V}\right)^{K_{0}^{\prime}} - 1\right]
\end{equation}
The bulk modulus K$_{0}$, which characterizes the stiffness of the
solid, was obtained below and above the structural transitions,
whereas the derivative of the bulk modulus was fixed at
K$_{0}^{'}$=4 i.e. the typical value for crystals with nearly
isotropic compression \cite{loa}. The best-fit curves obtained
using the Birch-Murnaghan are in good agreement with the
experimental results, as shown in Fig. \ref{FIG5}. The best-fit
values for K$_{0}$ for all the samples obtained in the low- and
high-pressure phases are shown in Tab. \ref{TAB1}, where a summary
of the main results discussed up to now is also reported.
\begin{table}[h]
  \centering
\begin{tabular}{|c|c|c|c|}
\hline
\multicolumn{4}{|c|}{LOW-PRESSURE PHASE}\\[1ex]
\hline
sample & phase & Orbital ordering & K$_{0}$\\
  \hline
  LG20 & \emph{Pbnm} O'& \emph{d}-type & 153 GPa\\
  LG30 & \emph{Pbnm} O'& \emph{d}-type & 160 GPa \\
  LG60 & \emph{Pbnm} O & no & 188 GPa \\[1ex]
  \hline
\multicolumn{4}{|c|}{HIGH-PRESSURE PHASE}\\
   \hline
    sample & phase & Orbital ordering & K$_{0}$\\
    \hline
  LG20 &  \emph{Imma}& ? & 209 GPa \\
  LG30 &  \emph{R$\bar{3}$c} & no &  210 GPa \\
  LG60 &  \emph{R$\bar{3}$c} & no &  175 GPa \\
  \hline
\end{tabular}
\caption{K$_{0}$ values obtained by Birch-Murnaghan fit and the
main results}\label{TAB1}
\end{table}
As to the low-pressure phases, we notice that the K$_{0}$ value of
the orbital disordered LG60 is about 20$\%$ larger than those of
the orbital ordered LG20 and LG30. This result suggests that the
stiffness of the solid increases in the undistorted and orbital
disordered state. Bearing in mind this suggestion, some
conjectures can be drawn about orbital ordering in the
high-pressure phases. Indeed, we notice that  the LG20 and the
LG30 K$_{0}$ values remarkably increase (about 30 $\%$) on
entering the \emph{Imma} and the \emph{R$\bar{3}$c} phases and get
closer to the undistorted orbital disordered LG60 low-pressure
value. These arguments may sustain the  hypothesis that a specific
orbital order configuration does not occur in the LG20 \emph{Imma}
phase as well as in the LG30 \emph{R$\bar{3}$c} one. On the other
hand, the LG60 bulk modulus shows a different pressure behavior as
it slightly decreases at high pressure. This can be explained,
taking into account that the LG60 sample is undistorted and
orbital disordered in both low- and high-pressure phases.

\begin{figure}[h!]
  \includegraphics[width=9 cm]{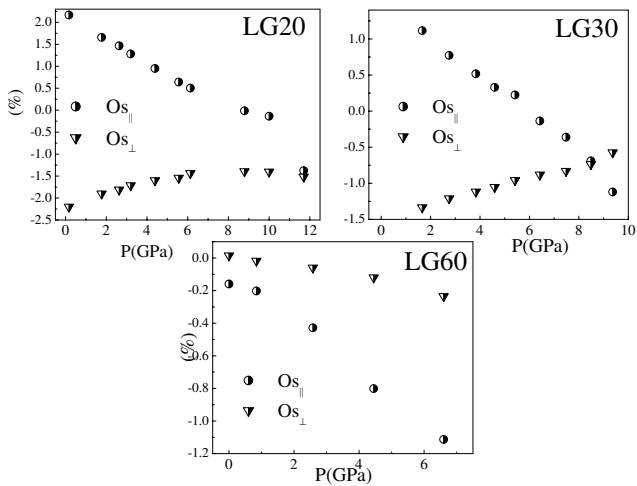}\\
  \caption{Pressure dependence of orthorhombic strain in the \emph{ab} plane (dots) and along the
   \emph{c} axis (diamonds).}\label{FIG6}
\end{figure}

In order to gain a further insight on the pressure dependence of
the structural properties of the investigated sample, the
orthorhombic strains in the \emph{ab} plane, Os$_{\|}= 2(a- c)/(a
+ c)$, and along the \emph{c} axis with respect to the \emph{ab}
plane, Os$_{\bot} = 2(b/p(2) - (a + c)/2)/(b/p(2) + (a + c)/2)$,
have been calculate and reported in Fig.\ref{FIG6} for each
sample.
These parameter represent the deviation of the structure from an
ideal cubic cell and obviously reflect the lattice parameter
behaviors. Indeed, the distorted LG20 and LG30 compounds show a
similar pressure dependence, different from the LG60. As expected
Os$_{\|}$ is larger (2$\%$) in the more distorted compound (LG20),
decreases to down to zero (a = b) at around 8-10 GPa, where the
\emph{a}-\emph{b} crossing is observed, and becomes negative at
larger pressure. A similar behavior is observed in the LG30
Os$_{\|}$ pressure dependence, but Os$_{\|}$ becomes negative
earlier (6-7 GPa) in this compound than in the less doped one.
Opposite pressure behavior is observed for the out of plane
orthorhombic distortion Os$_{\bot}$ in both the LG20 and LG30
samples. At ambient pressure, it is negative and diminishes in
absolute values with increasing pressure. Therefore, both the
Os$_{\|}$ and the Os$_{\bot}$ pressure trends reveal that the
application of hydrostatic pressure favors the system
symmetrization in the LG20 and LG30 compounds. On the other hand,
the strain pressure dependence suggests that pressure is not able
to completely remove the \emph{JT} octahedra distortion. Indeed,
in the LG20 and LG30 compounds the structure cannot retain in the
more regular phase (Os$_{\|}$=0, a=b), and the orthorhombic
distortion increases in absolute value at higher pressures up to
the structural transitions. In the LG60 sample, Os$_{\|}$ and the
Os$_{\bot}$ have similar pressure dependence. They started from
around 0 (\emph{Pbnm} O type structure with regular octahedra) and
becomes negative as the pressure increases. Both the Os$_{\bot}$
and the (Os$_{\|}$ increase in absolute value with pressure. This
phenomenon can be connected to a pressure induced octahedra
rearrangement which leads to the structural transition at 7 GPa.
\subsection{High pressure Raman spectroscopy}
\begin{figure}[h!]
  \includegraphics[width=8 cm]{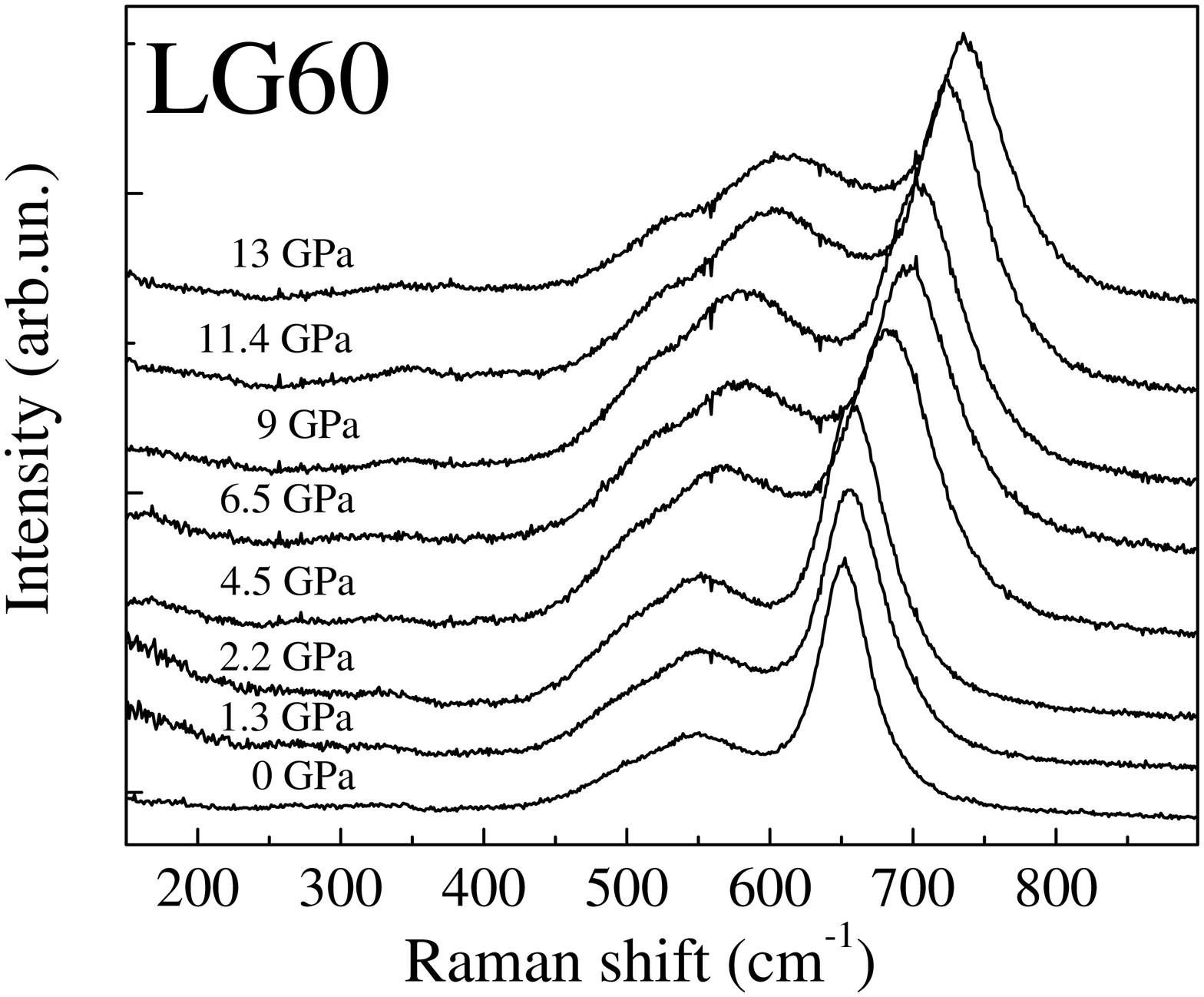}\\
  \caption{Raman spectra of LG60 at selected pressure. }\label{FIG7}
\end{figure}

\begin{figure}[h!]
  \includegraphics[width=8 cm]{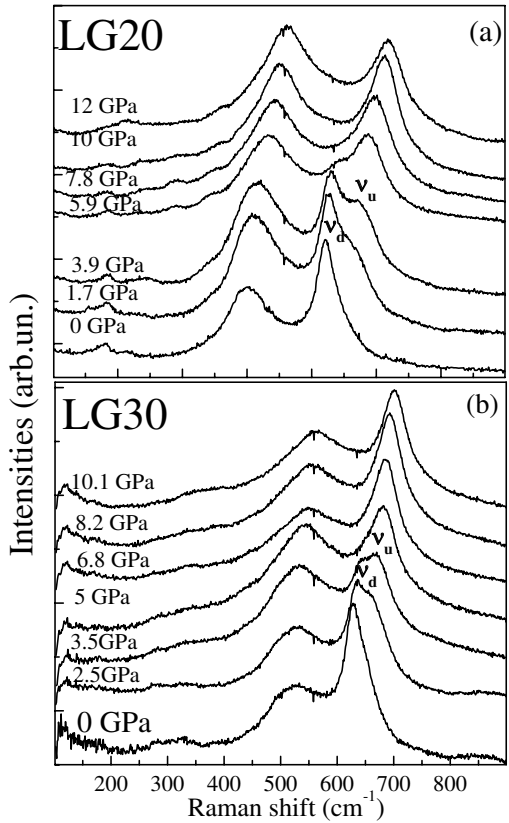}\\
  \caption{Raman spectra of LG20 panel (a) and LG30 panel (b) at selected pressure. }\label{FIG8}
\end{figure}

\begin{figure}[h!]
  \includegraphics[width= 7cm]{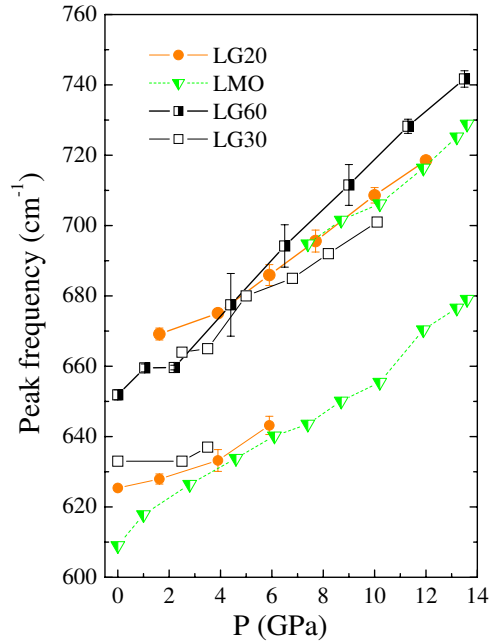}\\
  \caption{Pressure dependence of the$\nu_{d}$ and $\nu_{u}$
stretching frequencies of LaMnO$_{3}$ (Ref. \onlinecite{loa}),
LG20 and LG30 compared with that of the stretching frequencies of
the LG60.}\label{FIG9}
\end{figure}

Raman spectra of LG20 (Ref.\onlinecite{noigallati}), LG30 (present
results) and LG60 (Ref.\onlinecite{noigallati}) are shown in
Fig.\ref{FIG7} and Fig.\ref{FIG8} at selected pressures. At
ambient pressure, all the samples show two main phonon bands:  a
peak centered at around 650 cm$^{-1}$ ($\nu_{d}$) ascribed to
MnO$_{6}$ octahedra stretching mode which has been demonstrated to
be the the most sensitive to the extent \emph{JT} distortion
\cite{dho}, and a peak centered at 500 cm$^{-1}$ mostly ascribed
to bending modes \cite{noigallati}.
Looking at Fig.\ref{FIG7}, a part from the expected hardening of
the phonon frequencies induced by the lattice compression, no
dramatic variation of the spectral shape is observed in the Raman
spectra of LG60. In particular the Raman response does not appear
to be sensitive to the orthorhombic to rhombohedral transition
evidenced by present x-ray diffraction data between 6 and 8 GPa
and discussed in the previous section. This is not surprising
since the Raman spectrum of manganites is mainly due to octahedral
modes which are weakly affected by the slightly different spatial
arrangement of the octahedra in the two structures. An
experimental confirm of the weak sensitivity of the Raman response
to the orthorhombic/rhombohedral transition can be found in
previous papers where the Raman  spectra of both orthorhombic and
rhombohedral LaMnO$_{3}$ are shown \cite{Iliev1, iliev2}. The
spectra of this compound in the two structural phases are indeed
very similar, only a peak broadening and a reduction of  intensity
is observed in Raman spectrum of the rhombohedral LaMnO$_{3}$
sample with respect to the orthorhombic one \cite{Iliev1, iliev2}.
Bearing in mind this last result, we note that the Raman spectrum
of the LG60 shows a rather abrupt peak broadening at 4.5 GPa (see
Fig.\ref{FIG7}) that is at about the same pressure at which the
x-ray diffraction patterns show a broadening of the Bragg peak at
15$^{\circ}$ whose splitting marks the structural transition.
The Raman spectrum of the distorted LG30 sample is shown at
selected pressure in Fig. \ref{FIG8} (a). The pressure dependence
of the LG30 Raman spectrum is rather similar to that observed for
the most-distorted LG20 (Ref.\onlinecite{noigallati}) shown in
Fig. \ref{FIG8} (b) for sake of comparison, whereas both are quite
different from the pressure dependence of the LG60 undistorted
compound (see Fig.\ref{FIG7}). We notice that Raman spectra of
both LG30 and LG20 have been collected in orthorhombic
low-pressure phase (see the previous x-ray diffraction section)
thus allowing to follow the octahedral symmetrization process.
Within this context, the most important spectral feature is the
appearance of a high-frequency shoulder ($\nu_{u}$) of the
stretching peak $\nu_{d}$ in the Raman spectra of both LG20 and
LG30 on applying a moderate pressure. On further increasing the
pressure, a progressive transfer of spectral weight from the
pristine stretching peak $\nu_{d}$, to the new pressure-activated
peak, $\nu_{u}$, occurs. This pressure-induced process
extinguishes at P=8 GPa for LG30 and at P=10 GPa for LG20 when the
$\nu_{u}$ peak only is observed. Actually the same behavior under
applied pressure (appearance of a new pressure-activated $\nu_{u}$
peak and transfer of spectral weight), has been previously
observed for the Raman spectrum of the parent compound LaMnO$_{3}$
\cite{loa}. In this case the pristine stretching peak ($\nu_{d}$)
still survive, albeit with a strongly reduced intensity, at the
highest measured pressure (16 GPa) \cite{loa}. Nevertheless an
extrapolation of the experimental data at higher pressure shows
that the spectral transfer is completed at around 18 GPa
\cite{loa}. This peculiar behavior has been interpreted as a
signature of the octahedral symmetrization process
\cite{loa,noigallati}. The Raman spectra of LaMnO$_{3}$, LG20, and
LG30 at ambient pressure show a single $\nu_{u}$ peak ascribed, as
mentioned above, to the octahedral stretching mode of distorted
octahedra ($\nu_{d}$). The appearance of the new $\nu_{u}$ peak on
applying the pressure is thus ascribed to the onset of domain of
undistorted or less-distorted octahedra with a slightly higher
stretching frequency $\nu_{u}$. The pressure induced transfer of
spectral weight  from $\nu_{d}$ to $\nu_{u}$ is thus simply the
spectral evidence of a pressure-induced conversion of distorted
octahedra into regular ones \cite{noigallati}.\\It is particularly
interesting to compare the pressure behavior of the $\nu_{d}$ to
$\nu_{u}$ stretching peaks obtained  for the three samples
reported in the present paper and for the LaMnO$_{3}$ \cite{loa}
as shown in Fig. \ref{FIG9}. On applying pressure, the new
$\nu_{u}$ peak in the LaMnO$_{3}$, LG20, and LG30 spectrum appears
at about the same frequency of the stretching phonon of LG$60$
and, after the completion of the spectral weight transfer the
highest frequency peaks of the three samples show about same
pressure dependence. Bearing in mind that the LG$60$ lattice
consists of regular octahedra only and that the onset and the
completion of the symmetrization process occurs at pressure which
decreases on decreasing the extent of \emph{JT} distortion (i.e on
going from LaMnO$_{3}$ to LG20 to LG30) the comparative analysis
of the pressure behavior of the $\nu_{d}$ to $\nu_{u}$ peaks
strongly support their assignments to stretching modes related to
distorted and undistorted (less-distorted) octahedra
\cite{noigallati}.
The simultaneous occurrence of both the $\nu_{d}$ and $\nu_{u}$
peaks within certain pressure ranges depicts a phase separation
scenario, where two kinds of MnO$_{6}$ octahedra are
simultaneously present. At intermediate pressures, the coexistence
of domains of distorted and more regular octahedra can be clearly
claimed for both the LG20 and LG30 compound and this finding can
be obviously extended to the LaMnO$_3$ case previously
investigated \cite{loa}. The analysis of the Raman measurements
thus confirms the results discussed in the previous x-ray
diffraction section and adds further and more detailed information
about the symmetrization process, not revealed by the high
pressure diffraction measurements.
\section{Discussion and conclusion}
\begin{figure}
  \includegraphics[width=10 cm]{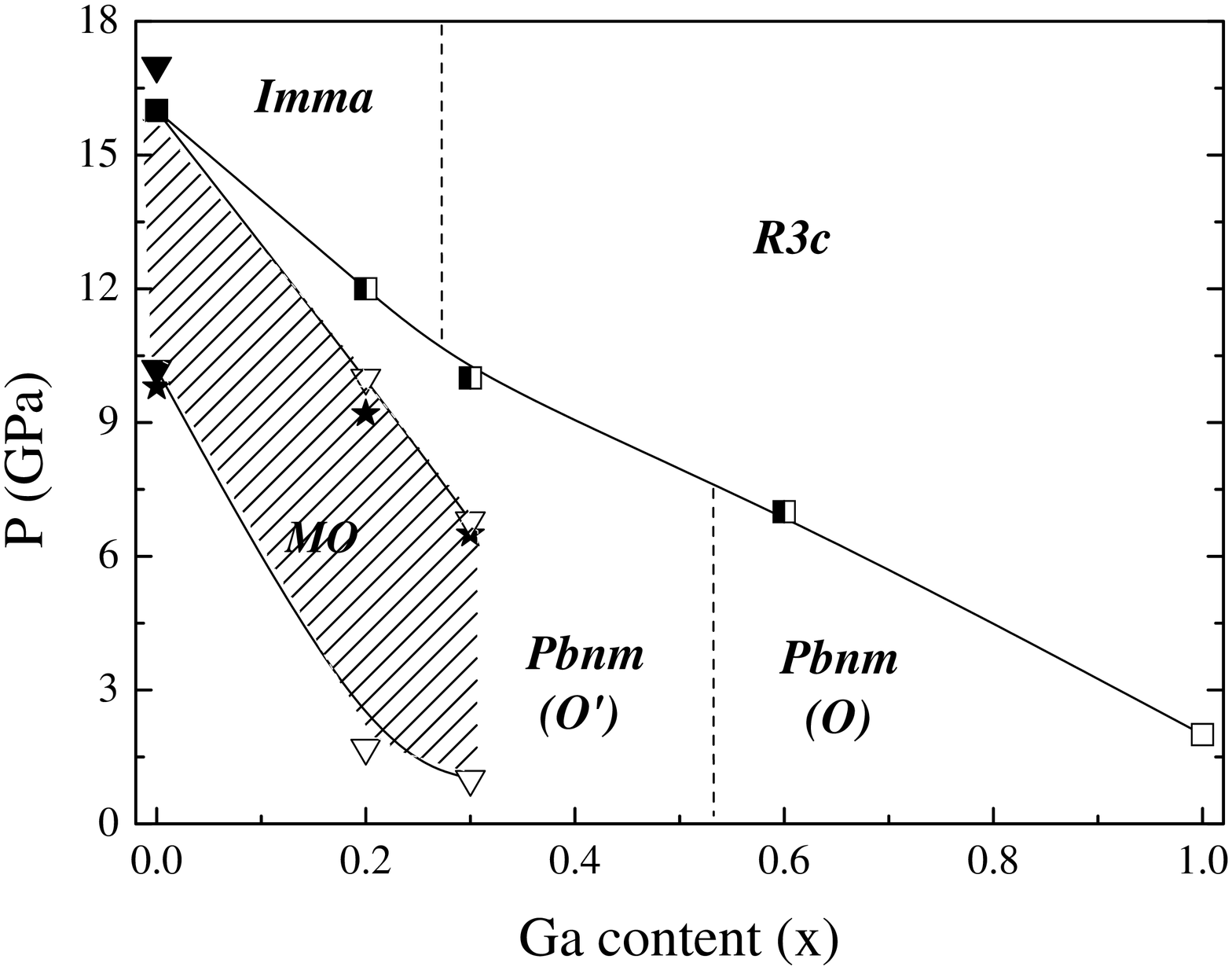}\\
  \caption{P vs Ga-concentration phase diagram.
Black square and triangles: X-ray and Raman results on LaMnO$_{3}$
from ref. \cite{loa}; white square: X-ray results on LaGaO$_{3}$
from ref. \cite{LaGaO3}; black and white squares and white
triangles: X-ray and Raman present data, respectively. The stars
indicate the pressure where the \emph{a}-\emph{b} lattice
parameter crossing occurs. MO= Mixed Octahedra. }\label{FIG10}
\end{figure}

The analysis of the high-pressure X-ray diffraction data shows
that all the three investigated samples undergo a pressure induced
transition from the orthorhombic \emph{Pbnm} phase towards
structures with higher symmetry. The three samples at ambient
pressure show a different extent of octahedral \emph{JT}
distortion which decreases going from LG20, to LG30, and to LG60,
where the octahedra are are not distorted. The transition pressure
to the higher symmetry phase progressively decreases on reducing
the extent of \emph{JT }distortion being P=12 GPa for LG20, P=10
GPa for LG30, and P=7 GPa for LG60. The high-pressure phase is a
more symmetric orthorhombic structure (\emph{Imma}) for the most
distorted sample, LG20, whereas LG30 and LG60 enter the
rhombohedral (\emph{R$\overline{3}$c}) phase. This results clearly
indicate a key role of the \emph{JT} distortion in stabilizing the
manganite structures. Moreover, the pressure dependence of the
structural parameters obtained from the two samples (LG20 and
LG30), distorted at ambient pressure, clearly demonstrates that
the most important structural effect of pressure is that of
reducing the octahedral distortion. The last finding is confirmed
by the analysis of the Raman data. In particular a pressure
induced transition from a distorted octahedral configuration to a
more regular one at high pressure is well evident in the observed
pressure dependence of the Raman-active stretching modes of LG20
and LG30. Moreover the peculiar pressure behavior shown by the
Raman spectra of LG20 and LG30 depicts a phase separation
scenario. On applying a moderate pressure a second stretching peak
associated to the onset of domains of less distorted octahedra
appears in the Raman spectra. On further increasing the pressure,
the number and/or the extension of the distorted octahedra domains
is progressively reduced down to zero and after the completion of
this process the structural transitions to more symmetric phases
occur. The simultaneous presence of two stretching peaks
associated to distorted and less distorted octahedra at
intermediate pressure is thus the clear signature of a phase
separation regime. Basing on the present results and using those
reported in Ref. \cite{loa} and in ref.\cite{meneghini} for
LaMnO$_{3}$, in Ref. \cite{gapressione} for LaGaO$_{3}$ and those
for from Ref. \cite{neutroni} for ambient pressure data, the
pressure vs. Ga-concentration phase diagram shown in Fig.
\ref{FIG10}, can be drawn. From the figure it is well evident that
the transition pressure to the higher symmetry phases
(orthorhombic \emph{Imma} and rhombohedral
\emph{R$\overline{3}$c}) decreases as the Ga content increases.
Therefore, the more symmetric phases appear to be promoted by the
octahedra symmetrization induced by the Ga content. The region,
where the symmetrization process occurs through the onset of
domains of less-distorted octahedra as obtained by present
measurements and from those reported in Ref. \cite{loa} is also
shown in Fig. \ref{FIG10}. In the same figure, the pressures at
which the \emph{a}-\emph{b} crossing for LG20 and LG30 (present
data) and for LaMnO$_{3}$ (Ref. \cite{loa}) is observed, are also
reported. This behavior is clearly strictly connected to the
symmetrization process and it can be reckon as a general
indication of the onset of a structural instability in the
perovskite structure of manganites which anticipates a structural
transition. This result thus suggest that a pressure induced
transition of the system from the orthorhombic to the cubic
structure appears to be very unlikely.\\
Although a pressure induced symmetrization process in LG20 and
LG30 as well as in LaMnO$_{3}$ (Ref. \onlinecite{loa}) is clearly
revealed by both x-ray diffraction and Raman measurements the
complete suppression of the \emph{JT} distortion in the
high-pressure phases cannot be claimed. Indeed, the
pressure-driven symmetrization process certainly suppresses the
\emph{coherent} \emph{JT} distortion in the higher symmetry phases
\emph{Imma} (LG20) and \emph{R$\overline{3}$c} (LG30) but the
occurrence of incoherent and/or dynamic MnO$_{6}$ \emph{JT}
distortion cannot be excluded. Finally indications on the orbital
ordered state in the high pressure phases can be inferred from the
analysis of the pressure dependence of the lattice parameters and
cell volumes. Above the structural transition, the occurrence of
orbital disordered state can be claimed for the rhombohedral LG30
sample, whereas in the orthorhombic \emph{Imma} LG20 the
persistence of some kind of orbital orientation cannot be ruled
out, but the occurrence of a specific orbital configuration seems
to be able to bring into question by the bulk modulus values
obtained with a Birch-Murnaghan curve. In conclusion the present
measurements provide a wide characterization of the pressure
behavior of three representative compounds belonging to the
LaMn$_{1-x}$Ga$_x$O$_3$ family and allow to draw a first sketch of
the P vs. Ga-content phase diagram. The obtained  results provide
also several general structural indications which can be exploited
in studying the effect of the chemical pressure in other manganite
systems at ambient pressure. Although open questions about the
effect of lattice compression on the persistence of some kind of
octahedral distortions and on the orbital ordering still remain,
the present data represent a fundamental base for extending to
this family of compounds the advanced theoretical studies carried
out on the parent LaMnO$_3$. Finally, to gain a deeper insight on
the effect of lattice compression, high-pressure neutron studies
on these compound would be necessary to investigate the effects of
lattice compression on the magnetic ordering.

\end{document}